\documentclass[journal,10pt,letterpaper,final,twocolumn]{IEEEtran}%
\usepackage{amsmath}
\usepackage{amsfonts}
\usepackage{cite}
\usepackage{amssymb}
\usepackage{mathrsfs}
\usepackage{graphicx}
\usepackage[usenames,dvipsnames]{color}
\usepackage{epstopdf}
\usepackage[all]{xy}
\usepackage[]{mcode}
\usepackage{algorithm}
\usepackage{algorithmic}
\usepackage{subcaption}
\usepackage{array}

\providecommand{\U}[1]{\protect\rule{.1in}{.1in}}
\pdfoutput=1
\providecommand{\U}[1]{\protect\rule{.1in}{.1in}}

\newenvironment{definition}[1][Definition]{\begin{trivlist}
\item[\hskip \labelsep {\bfseries #1}]}{\end{trivlist}}

\newcommand{\qed}{\nobreak \ifvmode \relax \else
      \ifdim\lastskip<1.5em \hskip-\lastskip
      \hskip1.5em plus0em minus0.5em \fi \nobreak
      \vrule height0.75em width0.5em depth0.25em\fi}

\begin{document}

\title{\LARGE {Energy-Efficient Fixed-Gain AF Relay Assisted OFDM  \\with Index Modulation}}
\author{Jiusi Zhou, \textit{Student Member, IEEE}, Shuping Dang, \textit{Member, IEEE}, Basem Shihada, \textit{Senior Member, IEEE}, and \\ Mohamed-Slim Alouini, \textit{Fellow, IEEE} 
  \thanks{J. Zhou, S. Dang, B. Shihada, and M.-S. Alouini are with Computer, Electrical and Mathematical Science and Engineering Division, King Abdullah University of Science and Technology (KAUST), Thuwal 23955-6900, Kingdom of Saudi Arabia (e-mail: $\{$jiusi.zhou, shuping.dang, basem.shihada, slim.alouini$\}$@kaust.edu.sa).}
}

\maketitle

\begin{abstract}
To broaden the application scenario and reduce energy consumption, we propose an energy-efficient fixed-gain (FG) amplify-and-forward (AF) relay assisted orthogonal frequency-division multiplexing with index modulation (OFDM-IM) scheme in this letter. The proposed system needs neither instantaneous channel state information (CSI) nor complicated processing at the relay node. It operates based on the power allocation scheme that minimizes the sum of transmit power at both source and relay node, given an outage probability constraint. Through a series of problem transformation and simplification, we convert the original power allocation problem to its relaxed version and solve it using convex programming techniques. To reveal the computing efficiency of the proposed power allocation scheme, we analyze its computational complexity. Numerical simulations substantiate that the proposed optimization scheme has a neglectable loss compared with the brute force search, but the computational complexity can be considerably reduced. 
\end{abstract}


\section{Introduction}
\IEEEPARstart{O}{rthogonal} frequency-division multiplexing with index modulation (OFDM-IM) is a frequency-domain extension of spatial modulation (SM) and has been regarded as one of the most promising candidate modulation schemes for next generation networks \cite{b3,8315127}. Theoretical and numerical results have verified the spectrum-efficient nature of OFDM-IM \cite{7330022}. Recently, to improve the coverage and reliability of OFDM-IM systems, the relay assisted OFDM-IM system was first investigated through numerical results in \cite{8075970}. A number of works have been published following this groundbreaking work to evaluate relay assisted OFDM-IM, mainly incorporating decode-and-forward (DF) relaying protocols \cite{b5,b6,8405601,8891788,8917612}. However, the DF relaying is employed in most works, which requires the received signal to be decoded at the relay node by instantaneous channel state information (CSI), resulting in elevated system complexity. Despite that the DF relaying has superior performance compared to the amplify-and-forward (AF) relaying, due to the relay node's energy and complexity constraints, DF relay system may not be supported in practice. Therefore, researchers currently dedicate more efforts to AF relay assisted OFDM-IM \cite{b9,8730295,8612925}, which study the error and outage performance as well as the comparison between fixed-gain (FG) relaying protocol and variable-gain (VG) relaying protocol, while these studies to some extent omit the optimization of the system from the energy-efficient perspective.

For cooperative systems, utilizing energy as efficiently as possible plays a crucial role at the source and relay nodes \cite{dang2020should}. In this regard, there exist numerous energy-efficient optimization schemes suited for different cooperative applications. Convex programming with partial CSI provides a prospect to approach optimal power allocation for AF relay systems \cite{new1}. Meanwhile, the iterative algorithm is also proposed to maximize the system energy efficiency for relay selection \cite{new2}. To further reduce the computational complexity, the sub-optimal solution can also be regarded as asymptotically optimal in certain applications, which develops into a new means to optimize the energy efficiency of relay systems \cite{new3}.

To further improve the energy efficiency of FG AF relay assisted OFDM-IM systems, we propose an energy-efficient FG AF relay assisted OFDM-IM scheme, which is capable of reducing the total transmit power given an outage probability constraint. We also convert the original power allocation problem to a relaxed problem that can be solved by convex programming techniques within polynomial time. To reveal the computing efficiency of the proposed power allocation scheme, we analyze its computational complexity. Numerical simulations corroborate that the converted problem can yield comparable solutions with considerably reduced computational complexity compared to the brute force search.

\section{System Model}\label{sec2}
A classic three-node cooperative transmission model without direct transmission link is investigated in this letter, where $N$ subcarriers in the set $\mathcal{N}$ are utilized to convey information from source to destination via an FG AF relay node. Following the modulation rules specified in \cite{b2}, we regulate that $T$ out of $N$ subcarriers are chosen to be active and a subcarrier activation pattern (SAP) is thereby established via a subset of active subcarriers $\mathcal{T}(k)$. It can be easily derived that by such an SAP, an extra bit stream with length $B_S= \lfloor \log_2 \binom{N}{T} \rfloor$ can be transmitted, and ${B_M=T \log_2 M}$ is the length of bit stream modulated by conventional amplitude-phase modulation (APM) scheme on $T$ active subcarriers, where $M$ is termed the APM order. Here, we assume the incoming bit stream with length ${B=B_M+B_S}$ is equiprobable for simplicity. By $N$-point inverse fast Fourier transform (IFFT) with the insertion of sufficiently long cyclic prefix, independent OFDM block covering all $N$ subcarriers for transmission is yielded and can be written as $\mathbf{x}(k)=[x(m_1,1),x(m_2,2),\dots,x(m_N,N)]^T \in \mathcal{C}^{N\times1}$, where $ {[\cdot]^{T} }$ denotes the transpose operation of vectors/matrices and $1\leq m_n\leq M$ is the index of APM symbol carried on the $n$th subcarrier. Here, we employ $M$-ary phase shift keying ($M$-PSK) as the APM scheme owing to its preferable property of constant envelope, and stipulate that for active subcarriers we have ${x(m_n , n)x(m_n , n)^* = 1}$, and ${x(m_n , n) = 0}$ otherwise. 

Assuming perfect CSI{\footnote{Considering practical hardware constraints, we assume that instantaneous and statistical CSI is perfectly known at the destination and the relay node, respectively. This configuration is different from those in the previous works investigating the DF relay and VG AF relay assisted OFDM-IM systems that also assume perfect instantaneous CSI at the relay node(s).}} and synchronization in the time and frequency domains, and neglecting the node mobility, we further suppose that all channels keep invariant within one complete transmission interval. Further assuming a Rayleigh fading environment, the end-to-end received OFDM block contaminated by noise terms can be expressed as \cite{b10}
\begin{align}  &{\mathbf{y}(k)} = [y(m_1,1),y(m_2,2),\dots,y(m_N,N)]^T \notag\\
&= \sqrt{\frac{P_r}{T}} \sqrt{\frac{P_t}{T}} {\mathbf{H}_1 \mathbf{H}_2 \mathbf{x}(k)} + \sqrt{\frac{P_r}{T}}  {\mathbf{H}_2 \mathbf{w}_1+\mathbf{w}_2} ,
\label{eq:Pn}\end{align}
where $P_r$ is the transmit power of the relay node, which can be regarded as the fixed amplification gain and is controllable at the FG AF relay node according to the statistical CSI; $P_t$ is the transmit power of the source; $P_r$ and $P_t$ are uniformly allocated over $T$ active subcarriers; $\mathbf{w}_{i}$ is the complex additive white Gaussian noise (AWGN) vector at the receiving side of the $i$th hop with average noise power $\eta_i = \mathbb{E}\{\mathbf{w}_i ^{2}\}$, where $\mathbb{E}\{\cdot\}$ denotes the expected value of the random variable enclosed; $\mathbf{H}_{i}=\operatorname{diag}\left\{h_{i}(1), h_{i}(2), \ldots, h_{i}(N)\right\}$ is the diagonal channel state matrix (DCSM) of the $i$th hop. Because the Rayleigh fading environment is assumed, we can have the probability density function (PDF) and the cumulative distribution function (CDF) of the channel power gain (CPG) $G_{i}(n)=\left|h_{i}(n)\right|^{2}$ to be $f_{i}(\xi)=\exp \left(-\xi / \mu_{i}\right) / \mu_{i}$ and $F_{i}(\xi)=1-\exp \left(-\xi / \mu_{i}\right)$, where $\mu_{i}$ is the average CPG of the $i$th hop, which captures the average channel quality in the $i$th hop and is dependent on per-hop distance, large-scale fading, and terrain, etc. As a consequence of the assumption of independent subcarriers, $G_i(1), G_i(2),\dots, G_i(N)$ regarding all subcarriers are supposed to be independent and identically distributed (i.i.d.). That is, there is no frequency and cross-hop correlation.

According to (1) and the assumptions of transmission channel given above, the independent end-to-end signal-to-noise ratio (SNR) regarding an arbitrary active subcarrier is expressed as follows:
\begin{equation} \label{eq3}\small
\gamma(k,n)= \frac{P_t P_r G_1(n) G_2(n) }{{T}{P_r} G_2(n)\eta_1 +  {T}^2\eta_2 },~~\forall~n \in \mathcal{T}(k) .
\end{equation}

\section{Formulation of Outage Probability} \label{sec3}
Since OFDM-IM is different from classic OFDM and the outage probability is an important indicator of system reliability that is adopted as an optimization constraint in this letter, we briefly analyze the outage performance of FG AF relay assisted OFDM-IM systems in this section. First, we define the outage event of an OFDM transmission block as follows: 

\begin{definition} \label{def:1}
An outage event of an OFDM transmission block happens if the end-to-end SNR of any active subcarrier is lower than a preset outage threshold $s$.   
\end{definition}

By this definition of outage event, we can write the average outage probability that characterizes the end-to-end reliability of the proposed relay assisted OFDM-IM system to be $P_{o}(s)=\underset{k}{\mathbb{E}}\left\{\mathbb{P}\left\{\bigcup_{n \in \mathcal{T}(k)}\{\gamma(k, n)<s\}\right\}\right\}$, where $\mathbb{P}\{\cdot\}$ represents the probability of the random event enclosed. Because of the union relation linking all subcarrier-wise outage events, we can resort to the subcarrier independence and the inclusion-exclusion principle to derive $P_{o}(s)=\underset{k}{\mathbb{E}}\left\{1-\prod_{n \in \mathcal{T}(k)}(1-\mathbb{P}\{\gamma(k, n)<s\})\right\}$. Since the fading channels of all subcarriers are statistically independent, we have $\Phi(s)=\mathbb{P}\left\{\gamma\left(k, n_{1}\right)<s\right\}=\mathbb{P}\left\{\gamma\left(k, n_{2}\right)<s\right\} \quad=\quad \cdots \quad=\quad \mathbb{P}\left\{\gamma\left(k, n_{T}\right)<s\right\}$, $n_{1}, n_{2}, \ldots, n_{T} \quad \in \mathcal{T}(k)$, which further simplifies $P_{o}(s)$ to be
\begin{align} \label{eq6}
&P_{o}(s) =\sum_{k=1}^{2^{B_{S}}}\{[1-\left.\left.(1-\Phi(s))^{T}\right] / 2^{B_{S}}\right\}=1-(1-\Phi(s))^{T},
\end{align}
where $\Phi(s)$ is the subcarrier-wise outage probability. That is, the average outage probability can be regarded as independent from SAP and the total number of subcarriers $N$. Instead, the average outage probability of the proposed system is dominated by the number of active subcarriers $T$. Subsequently, we can determine $\Phi(s)$ by \cite{b10}
\begin{equation} \label{eq7}\small
\Phi(s) =1-{2}{T} \sqrt{\frac{s \eta_{2}}{\mu_{1} \mu_{2} {P_t} {P_r}}} \exp \left(-\frac{s{T} \eta_{1}}{\mu_{1} {P_t}}\right) K_{1}\left({2}{T}\sqrt{\frac{s \eta_{2}}{\mu_{1} \mu_{2} {P_t} {P_r}}}\right),
\end{equation}
where $K_v(\cdot)$ denotes the $v$th-order modified Bessel function of the second kind. Substituting (\ref{eq7}) into (\ref{eq6}) yields the closed-form expression of the average outage probability for the relay assisted OFDM-IM systems utilizing FG AF relaying.

\section{Optimization Problem Analysis} \label{sec4}

For energy-efficient communication systems, the total power utilized should be minimized, while maintaining outage probability $P_{o}(s)$ below a predetermined constraint $\Psi_{th}$. Meanwhile, the power supply is also constrained by the practical hardware on each node. $P_t^{\max}$ and $P_r^{\max}$ specify the maximum power available at the source and the relay, respectively.  As a result, the optimization problem can be formulated as

 \begin{equation}\small
 \begin{split}
 \mathrm{minimize}&~~~~P_t+P_r\\
 \mathrm{s.t.}&~~~~P_{o}(s)\leq \Psi_{th}\\
 &~~~~0 \leq P_{t} \leq P_{t}^{\max },~0 \leq P_{r} \leq P_{r}^{\max }
 \end{split}
 \end{equation}

Notice that for a non-negative random variable $X$, $\mathbb{E}\{X\}=\int_{0}^{\infty}\left(1-F_{X}(x)\right) \mathrm{d} x$ according to Lebesgue-Stieltjes integration \cite{b11}. Then the relation validates the equivalence between $\max \left\{\Phi(s)\right\}$ and $\min \left\{\mathbb{E}\left\{\gamma(k,n)\right\}\right\}$. Subsequently, we can determine $\mathbb{E}\left\{\gamma(k,n)\right\}$ by (\ref{eq:Pn}), where $\mathrm{Ei}\left(\cdot\right)$ returns the one-argument exponential integral defined as $\mathrm{Ei}\left(x\right) = \int_{-\infty}^{x} \frac{e^{t}}{t} d t$. According to Markov's inequality $\mathbb{P}(X\geq \nu)\leq \frac{\mathbb{E}(X)}{\nu}$, $\forall \nu > 0$, we have $\mathbb{E}\left\{\gamma(k,n)\right\} \geq {s}{\left({1-P_{o}(s)}\right)^{1/T}}$, which means that there is an equivalence between $\max \left\{P_{o}(s)\right\}$ and $\min \left\{\mathbb{E}\left\{\gamma(k,n)\right\}\right\}$. Therefore, the optimization constraint given by $P_{o}(s) \leq \Psi_{t h}$ can be equivalently transfered to $\mathbb{E}\left\{\gamma(k,n)\right\} \geq \gamma_{th}$. However, the specific numerical relationship between $P_{t h}$ and $\gamma_{th}$ is not straightforward to obtain. We present detailed steps to approximate the transformation between outage probability and average SNR in Appendix \ref{appendix:A}.

\begin{figure*}
\begin{equation}\label{eq:Pn}\small
\begin{split}
{\mathbb{E}\left\{\gamma(k,n)\right\}} &= 
{\int_{0}^{\infty} {2}{T} \sqrt{\frac{s \eta_{2}}{\mu_{1} \mu_{2} {P_t} {P_r}}} \exp \left(-\frac{s{T} \eta_{1}}{\mu_{1} {P_t}}\right)  K_{1}\left({2}{T}\sqrt{\frac{s \eta_{2}}{\mu_{1} \mu_{2} {P_t} {P_r}}}\right) \mathrm{d}s}={{\frac{\mu_{1} {P_t}}{{T} \eta_{1}}+\frac{\mu_1 \eta_2 P_t}{ \mu_2 \eta_1^{2} P_r} \exp\left(\frac{\eta_2}{\eta_1 \mu_2 P_r}\right)}\mathrm{Ei}\left(-\frac{\eta_2}{\eta_1 \mu_2 P_r}\right)}
\end{split}
\end{equation}
\rule{18cm}{0.01cm}
\end{figure*}

Meanwhile, because of mathematical intractability of $\mathbb{E}\left\{\gamma(k,n)\right\}$, we find an alternative to approximate $\mathbb{E}\left\{\gamma(k,n)\right\}$ for obtaining a tractable and sub-optimal solution. Let $\mu_i$ take place of $G_i(n)$, and then we obtain an alternative indicator of average end-to-end SNR $\widetilde{\gamma}(k,n)$ as $\widetilde{\gamma}(k,n)= \frac{P_t P_r  \mu_1 \mu_2 }{{T}{P_r} \mu_2 \eta_1 +  {T}^2\eta_2 }$. However, this indicator above is not close enough to $\mathbb{E}\left\{\gamma(k,n)\right\}$. In Appendix \ref{appendix:B}, we derive the gap denoted as $\delta$ between $\mathbb{E}\left\{\gamma(k,n)\right\}$ and $\widetilde{\gamma}(k,n)$, and we can further update the constraint as $\widetilde{\gamma}_{th} = \gamma_{th}+ \delta$. Based on the above updated constraints, we can transfer the original power allocation problem to a relaxed form, given the condition that $\widetilde{\gamma}(k,n)$ must be larger than the predetermined threshold $\widetilde{\gamma}_{th}$. Then, the relaxed optimization problem is given by

 \begin{equation}\small
 \begin{split}
 \mathrm{minimize}&~~~~P_t+P_r\\
 \mathrm{s.t.}&~~~~\widetilde{\gamma}(k,n) \geq \widetilde{\gamma}_{th}\\
 &~~~~0 \leq P_{t} \leq P_{t}^{\max },~0 \leq P_{r} \leq P_{r}^{\max }
 \end{split}
 \end{equation}

By analyzing the monotonicity and linearity, we can easily observe that the objective function and constraints are all convex in terms of $P_t$ and $P_r$. Additionally, because the objective function and $\widetilde{\gamma}(k,n)$ are monotonically increasing with the powers, the optimal solutions can be attained only if $\widetilde{\gamma}(k,n) = \widetilde{\gamma}_{th}$ . Since the other constraints are affine,  Slater's condition is satisfied, by which we can prove the strong duality of the formulated problem. As a result, we employ the Lagrange multiplier to derive the sub-optimal solutions and formulate the Lagrangian function as $\mathcal{L}\left(P_t, P_r,\rho \right)=P_{t}+P_{r}+\rho\left(\widetilde{\gamma}_{th}-\frac{P_t P_r  \mu_1 \mu_2 }{{T}{P_r} \mu_2 \eta_1 +  {T}^2\eta_2 }\right)$, where $\rho$ is the Lagrange multiplier. Taking the derivative of the Lagrangian function with respect to $P_t$ and $P_r$ and equating it to zero give
\begin{equation}\label{eq1314} \footnotesize
\begin{cases}
1-\frac{\rho P_r \mu_1 \mu_2 }{T^2 \eta_2 + T P_r\mu_2 \eta_1 }=0\\
1+\rho \left( \frac{T P_r P_t \mu_1 \mu_2^2 \eta_1}{\left(T^2 \eta_2+T P_r \mu_2 \eta_1\right)^2} - \frac{P_t \mu_1 \mu_2}{T^2 \eta_2 + T P_r \mu_2 \eta_1} \right)=0
\end{cases}.
\end{equation}
Also, by the Karush-Kuhn-Tucker (KKT) conditions, we can obtain
\begin{equation}\label{eq15}\footnotesize
\widetilde{\gamma}_{th}- \frac{P_t P_r  \mu_1 \mu_2 }{{T}{P_r} \mu_2 \eta_1 +  {T}^2\eta_2 }= 0.
\end{equation}
Solving (\ref{eq1314}) and (\ref{eq15}) yields the sub-optimal solutions as
\begin{equation}\label{eq161718}\footnotesize
\begin{cases}
P_{t}=\left[\frac{\rho T^2 \mu_1 \eta_2}{\mu_2 \left(\rho \mu_1 -T\eta_1\right)^2}\right]_{0}^{P_{t}^{\max }}\\
P_{r}=\left[\frac{T^2 \eta_2}{\mu_2 \left(\rho \mu_1 - T \eta_1\right)}\right]_{0}^{P_{r}^{\max }}\\
\rho = \frac{T}{\mu_2}\left(\eta_1+\sqrt{\frac{\mu_1 \eta_2}{\mu_2 \widetilde{\gamma}_{th} }}\right)
\end{cases},
\end{equation}
where 
\begin{equation}\footnotesize
\left[x\right]^{m}_{n}=\left\{\begin{array}{ll}
{m} & {x\geq m}, \\
{x} & {m > x \geq n},\\
{n} & {x < n}.
\end{array}~~~~~ (m>n)\right.
\end{equation}
The sub-optimal solutions derived above verify the potential that the proposed scheme can provide power allocation based on statistical CSI. Subsequently, we can resort to a simple iterative method presented in Algorithm \ref{algorithm1} to obtain a specific allocation strategy for arbitrarily system and channel configurations. Note that, although the scheme stipulated above is described in the context of a single subblock of subcarriers, it can be easily extended to the whole cooperative OFDM-IM system consisting of multiple subblocks by repeating the scheme by multiple times, because of the mutual independence among all subblocks \cite{b2}.

\begin{algorithm}[t!] 
 \caption{Algorithm for the Proposed Iterative Method.}
 \begin{algorithmic}[1]
 \renewcommand{\algorithmicrequire}{\textbf{Input:}}
 \renewcommand{\algorithmicensure}{\textbf{Output:}}
 \REQUIRE $T, \eta_1, \eta_2, \mu_1, \mu_2, {\gamma}_{th}, P_t^{\max}, P_r^{\max}$ 
 \ENSURE  $P_t, P_r$
  \\ \textbf{Initialisation:} $P_t = 0, P_r = 0, \epsilon = 10^{-4}, \omega = 1$;
  \WHILE {$\omega = 1$}
  \STATE $\delta \gets \frac{P_t P_r^2 T \eta_1 \mu_1 \mu_2^2}{\left(T P_r \eta_1 \mu_2 + T^2 \eta_2\right)^2}$;
  \STATE $\widetilde{\gamma}_{th} \gets {\gamma}_{th} + \delta $;
  \STATE $\rho \gets \frac{T}{\mu_2}\left(\eta_1+\sqrt{\frac{\mu_1 \eta_2}{\mu_2 \widetilde{\gamma}_{th} }}\right)$;
  \STATE $P_{t}^{*} \gets \left[\frac{\rho T^2 \mu_1 \eta_2}{\mu_2 \left(\rho \mu_1 -T\eta_1\right)^2}\right]_{0}^{P_t^{\max}}$;
  \STATE $P_{r}^{*} \gets \left[\frac{T^2 \eta_2}{\mu_2 \left(\rho \mu_1 - T \eta_1\right)}\right]_{0}^{P_r^{\max}}$;
  \IF {($\left|P_t-P_t^*\right|<\epsilon \And \left|P_t-P_t^*\right|<\epsilon$)}
  \STATE $\omega \gets 0$;
  \ELSE 
  \STATE $P_t \gets P_t^*$; $ P_r \gets P_r^*$;
  \ENDIF 
  \ENDWHILE
 \end{algorithmic} 
 \label{algorithm1}
 \end{algorithm}
 
\section{Computational Complexity Analysis}\label{secadd}

For comparison purposes, we analyze the computational complexities of the brute force search and the proposed scheme in this section. First, we set a uniform accuracy $\epsilon = 10^{-4}$, which is consistent with the precision specified in Algorithm \ref{algorithm1}. Based on the principle of the brute force search, we simplify the function seeking the optimized solution as $\mathcal{F}(\alpha)$ and denote the set of object data including $P_r$ and $P_t$ as $\alpha$. Obviously, this is a sub-linear coverage approaching problem aiming to ensure $\mathcal{F}\left(\alpha_{\kappa}\right)-\mathcal{F}^{*} \leq \frac{\theta}{\sqrt{\kappa}}$, where $\mathcal{F}^{*}$ is the optimal solution; $\kappa$ denotes the number of searching rounds; $\theta$ is set as a constant depending on hardware platform configurations and optimization requirements. Let $\frac{\theta}{\sqrt{\kappa}} \leq \epsilon$, and then we obtain $\kappa \geq \frac{\theta^{2}}{\epsilon^{2}}$, which leads to the computational complexity expressed as $\mathcal{O}\left(\frac{1}{\epsilon^{2}}\right)$.

As for the proposed power allocation scheme, it is in essence a linear coverage problem that can be expressed as $\left\|\alpha_{\kappa}-\alpha^{*}\right\| \leq \theta(1-q)^{j}$, where $\alpha^{*}$ is the converged results; the ratio $q$ is within the range of $(0,1)$ and $j$ denotes the iterative epochs. Similarly, we set the upper bound of the right side as $\theta(1-q)^{j} \leq \epsilon$. Then, solving it yields the result as $j \geq \frac{1}{q}\left(\log(\theta)+\log\left(\frac{1}{\epsilon}\right)\right)$, by which we can derive the computational complexity as $\mathcal{O}\left(\log\left(\frac{1}{\epsilon}\right)\right)$. This shows a significant superiority in terms of computing efficiency over the brute force searching that has exponential complexity. The traditional FG AF relay power allocation proposed in \cite{b10} provides an established gain as $\frac{P_t}{P_t \mu_1 + T \eta_2}$. For a given constraint of outage probability, the optimization strategy is to adapt the source and relay transmit power allocation in a distributed manner. As a result of the distributed nature, it frequently leads to power overflow at the source and/or the relay end, so that outage performance becomes unsatisfactory (this can be shown by the simulation results presented in the next section). Although the traditional method has much lower complexity given by $\mathcal{O}\left(1\right)$, it is inapplicable for reliability-critical cooperative networks. For clarity, we compare these three power allocation schemes by listing their computational complexities and precision requirements in Table \ref{table}.

\begin{table}  \small 
\caption{Computational complexities and precision requirements of different power allocation schemes.}
\begin{equation}\begin{array}{c|c|c}
\hline \text { Schemes } & \text { Complexity } & \text { Precision Requirement } \\
\hline \hline \text { Brute force Search } & \mathcal{O}\left(\frac{1}{\epsilon^{2}}\right) & \text { YES } \\
\hline \text { Proposed scheme } & \mathcal{O}\left(\log\left( \frac{1}{\epsilon}\right)\right) & \text { YES } \\
\hline \text { Traditional method } & \mathcal{O}\left(1\right) & \text { NO } \\
\hline
\end{array}\notag \end{equation}
\label{table}
\end{table}

\section{Numerical Results}\label{sec5}
\begin{figure}[t]
\centering 
\includegraphics[width =2.8in]{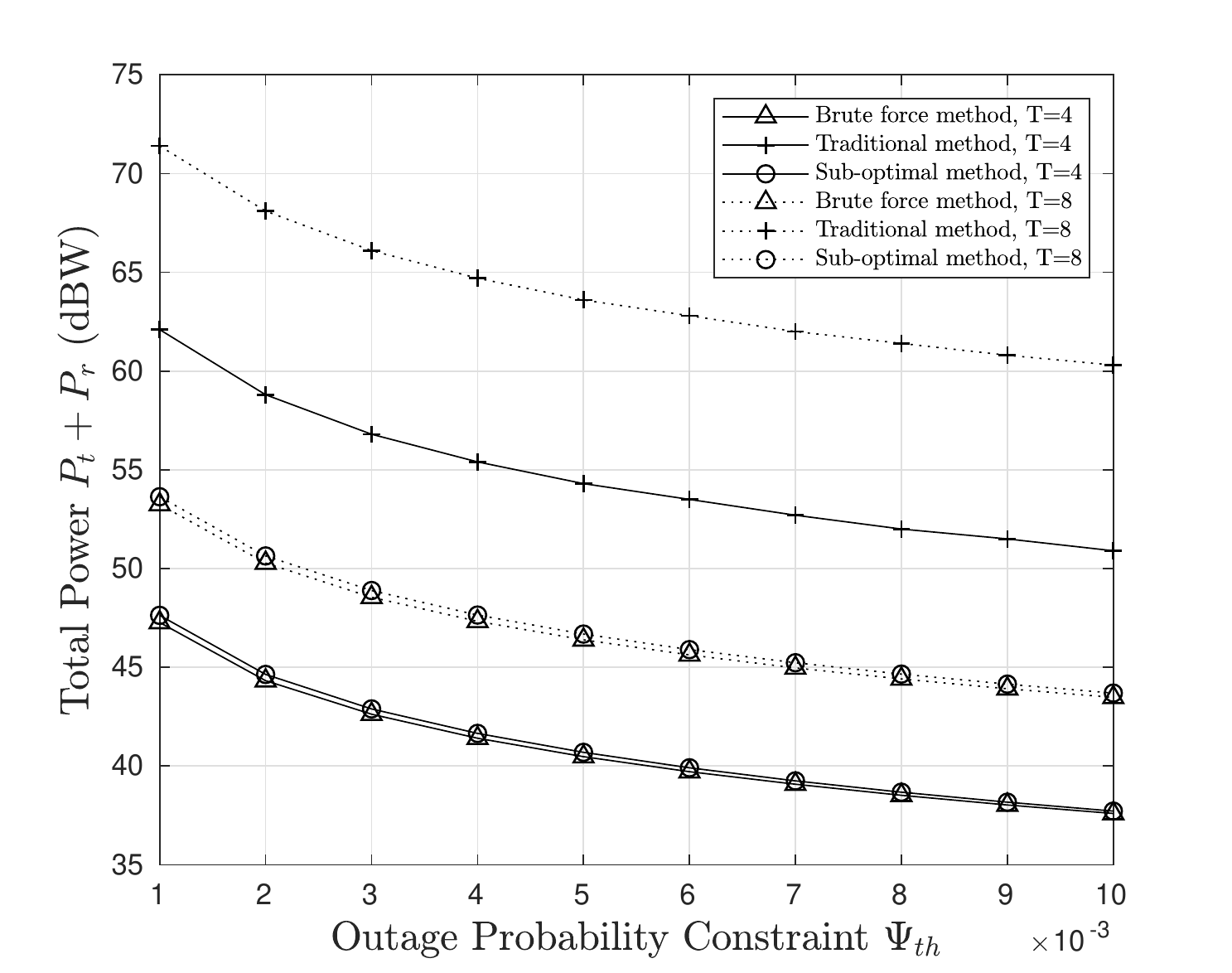}
\caption{Total transmit power vs. outage probability constraint, given $s=5$ dB.}
\label{fig:figure1label}
\end{figure}

\begin{figure}[t]
\centering 
\includegraphics[width =2.8in]{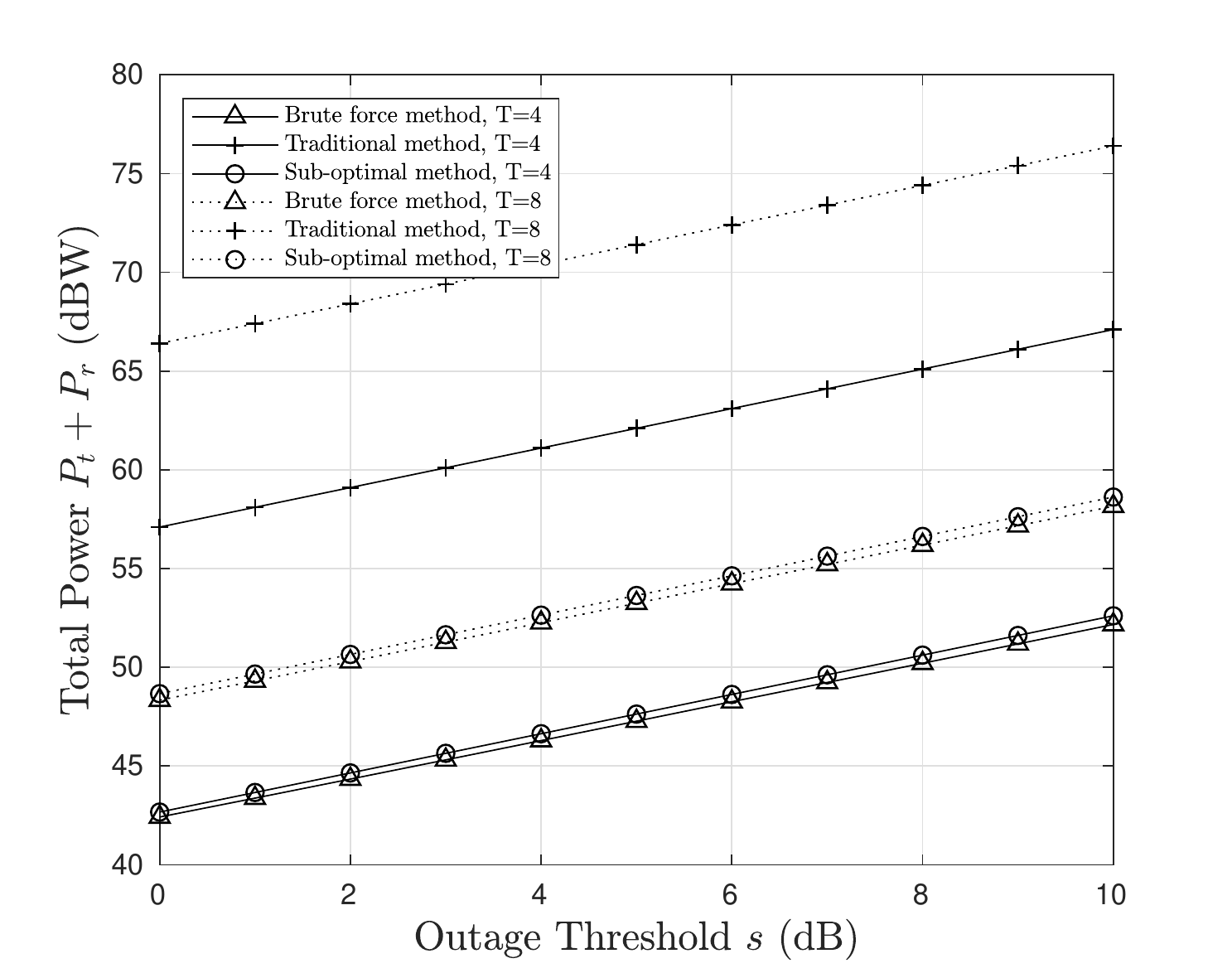}
\caption{Total transmit power vs. outage threshold, given $\Psi_{th} = 10^{-3}$.}
\label{fig:figure2label}
\end{figure}

To verify the effectiveness of the proposed optimization scheme, we carry out the numerical comparison between the sub-optimal solutions given in (\ref{eq161718}) with the optimal solutions generated by the brute force method as well as the traditional FG AF relay power allocation scheme proposed in \cite{b10}. In practice, the average power of thermal noise are always different for different subchannels over different hops, which are related to the equivalent ambient temperatures and bandwidths of the subcarrier. Meanwhile, the fading parameters can be different in both of the frequency and space domains as well. In this way, these fading parameters are capable of capturing the frequency selectivity and spatial heterogeneity of different hops. For these reasons, we set the simulation configurations as $\eta_1 = 1.3$, $\eta_2 = 1.1$, $\mu_1 = 1.3$, $\mu_2 = 1.5$, and $P_t^{\max}=P_r^{\max}=100$ $\rm{dBW}$. As for the brute force search, we utilize $10^{-4}$ $\rm{dBW}$ as the incremental basis to calculate the minimum total power that satisfies the conditions via a two-dimensional discrete grid corresponding to $\epsilon = 10^{-4}$. When we activate 4 and 8 subcarriers, the numerical comparison results are presented in Fig. \ref{fig:figure1label} and Fig. \ref{fig:figure2label}, which in particular focus on the relations between the total transmit power $P_t+P_r$ and the outage probability constraint $\Psi_{th}$ as well as the outage threshold $s$.

In Fig. \ref{fig:figure1label}, given outage threshold $s$ as $5$ dB, the performance curves reflect a declining trend of the minimum total transmit power $P_t+P_r$ with increasing outage probability constraint $\Psi_{th}$. The solid and dashed lines show the variations in total power when the number of active subcarriers is 4 and 8, respectively. The curve corresponding to the proposed scheme is in the midst of the other two, which almost overlaps the one of the brute force search, and is relatively separated from the one of the traditional scheme. Another group with 8 active subcarriers demonstrates analogous trends. Observing the numerical results, we can find that the proposed scheme possesses an obvious gain in the energy efficiency compared to the traditional scheme, and meanwhile, is close to the brute force search. Turning the target to the outage threshold $s$, Fig.  \ref{fig:figure2label} fixes the outage probability constraint to be $10^{-3}$. As the outage threshold $s$ raises, the total power $P_t+P_r$ is boosted. Meanwhile, the relations among different curves are similar as those shown in Fig.  \ref{fig:figure1label}.

From these two figures, we can easily summarize that the proposed power allocation strategy can achieve comparable performance as the optimal performance yielded by the brute force search. In addition, in comparison with the traditional FG AF relay power allocation scheme, our proposal is able to significantly reduce the total power required to achieve certain outage performance, which leads to a much higher energy efficiency and is thereby in line with the goal of green communications. Further taking the polynomial-time complexity nature into consideration, the effectiveness of the proposed power allocation scheme can be verified.

\section{Conclusion}\label{sec6}
In this letter, we proposed an energy-efficient FG AF relay assisted OFDM-IM scheme, aiming at reducing the total transmit power given an outage probability constraint. We investigated the formulated power allocation problem and relaxed it by some approximations. The relaxed problem can thus be solved by convex programming techniques within polynomial time and the yielded solutions are comparable to the solutions generated by the brute force search. The analysis of computational complexity was also provided to verify the computing advantage of the proposed power allocation scheme for real-time communications.

\appendices

\section{Objective Transformation between Outage Probability and Average SNR} \label{appendix:A}

In order to facilitate the following analysis, we define two parameters : $a = 2T \sqrt{\frac{\eta_2}{P_t P_r \mu_1 \mu_2}}$ and $b = \frac{T \eta_1}{P_t \mu_1}$. Accordingly, we obtain $P_{o}(s)= 1-\left(a\sqrt{s} \exp{\left(-bs\right)} K_1\left(a\sqrt{s}\right)\right)^T$ and ${\mathbb{E}\left\{\gamma(k,n)\right\}} = \frac{1}{b}+\frac{a^2}{4b^2}\exp{\left(\frac{a^2}{4b}\right)} \mathrm{Ei} \left(-\frac{a^2}{4b}\right)$. Then, according to the asymptotic formula from \cite{b12}: $K_{v}(x) \sim \frac{1}{2} \Gamma(v)\left(\frac{x}{2}\right)^{-v}$, when $v>0$ and $x \rightarrow 0$, where $\Gamma(\cdot)$ returns the complete gamma function, we can further approximate $P_o(s)$ to be
\begin{align} \label{eq22}
P_o(s) & \sim  
{1-\left(a\sqrt{s} \exp{\left(-bs\right)} \frac{1}{2} \Gamma(1)\left(\frac{a\sqrt{s}}{2}\right)^{-1}\right)^T}\notag\\
&={1-\left(\exp{\left(-bs\right)}\right)^T}<\Psi_{th}.
\end{align}
Consequently, we combine it with the asymptotic formula from \cite{b13} to be $e^{-x}\mathrm{Ei}\left(x\right) \geq -\log \left(1-\frac{1}{x}\right)$, and finally we have  
\begin{equation}\small
{\mathbb{E}\left\{\gamma(k,n)\right\}} \geq \frac{1}{b}\left(1-\frac{a^2}{4b}\log{\left(1+\frac{4b}{a^2}\right)}\right) .
\end{equation}

Given a defined function $f(x) = 1-x \log \left(1+\frac{1}{x}\right)$, it is obvious that $f(x)\rightarrow1$, when $x\rightarrow0$. Therefore, we can obtain the approximation of $\min \left\{ {\mathbb{E}\left\{\gamma(k,n)\right\}} \right\}$ as $\frac{1}{b}$. Meanwhile, by (\ref{eq22}), we can derive $b<\left. \log \left(\frac{1}{1-\Psi_{th}}\right) \middle / (sT) \right.$. Without loss of generality, we can simply let $\gamma_{th} = \left. sT \middle / \log \left(\frac{1}{1-\Psi_{th}}\right) \right.$.

\section{Derivation of the Gap between $\mathbb{E}\left\{\gamma(k,n)\right\}$ and $\widetilde{\gamma}(k,n)$}\label{appendix:B}

In order to derive the numerical value of the lower bound on average SNR $\mathbb{E}\left\{\gamma(k,n)\right\}$, we use Jensen gap to approach it, which is given by $\delta= \left|\mathbb{E}\left\{\gamma(k,n)\right\} - \widetilde{\gamma}(k,n)\right|$ \cite{b14}. According to (\ref{eq3}), we can express $\mathbb{E}\left\{\gamma(k,n)\right\}$ in an alternative form as $\mathbb{E}\left\{\gamma(k,n)\right\}= P_t P_r \mu_1 \mathbb{E}\left\{\left.{1}\middle/\left({T P_r \eta_1 + \frac{T^2 \eta_2}{G_2(n)}}\right)\right.\right\}$. To align with the form given above, we define an auxiliary function as $g(G_2(n))=\left.{1}\middle/\left({T P_r \eta_1 + \frac{T^2 \eta_2}{G_2(n)}}\right)\right.$ to facilitate the following analysis. We can  easily obtain the derivative of $g(G_2(n))$ with respect to $G_2(n)$ as $g^{\prime}(G_2(n))=\left.{T^2 \eta_2}\middle/{\left(T^2 \eta_2 + T P_r \eta_1 G_2(n)\right)^2}\right.$. According to the lemma in \cite{b14}, we obtain 
\begin{equation}\label{dsakdjk2}\small
\delta \leq \sup_{x} \left\{ \frac{g(x)-g(\mu_2)-g^{\prime}(\mu_2)(x-\mu_2)}{(x-\mu_2)^{2}} \right\} \operatorname{Var}[G_2(n)] P_t P_r \mu_1, 
\end{equation}
where $\operatorname{Var}[X]$ denotes variance of $X$. Finally, because of the monotone nature of (\ref{dsakdjk2}) in terms of $x$, we derive
\begin{equation}\small
\delta \leq \frac{P_t P_r^2 T \eta_1 \mu_1 \mu_2^2}{\left(T P_r \eta_1 \mu_2 + T^2 \eta_2\right)^2}.
\end{equation}

\bibliographystyle{IEEEtran}
\bibliography{bib}

\end{document}